\documentclass[letterpaper, 10 pt, conference]{ieeeconf}  

\IEEEoverridecommandlockouts                              
\usepackage{graphics} 
\usepackage{epsfig} 
\usepackage{amsmath} 
\usepackage{amssymb} 
\usepackage{cite}
\usepackage{algorithm}
\usepackage{algorithmic}
\usepackage{subfigure}
\usepackage{comment}

\title{\LARGE \bf
	Performance Quantification of a Nonlinear Model Predictive Controller by Parallel Monte Carlo Simulations of a Closed-loop System 
}

\author{Morten Wahlgreen Kaysfeld, Mario Zanon, John Bagterp J{\o}rgensen
	\thanks{*M. W. Kaysfeld and  J. B. J{\o}rgensen are with the Department of Applied Mathematics and Computer Science, Technical University of Denmark, DK-2800 Kgs. Lyngby, Denmark. Mario Zanon is with IMT School for Advanced Studies Lucca, IT-55100 Lucca, Italy. }
	\thanks{Corresponding author: J. B. J{\o}rgensen (E-mail: {\tt\small jbjo@dtu.dk}).}
}

\begin{document}
\maketitle
\thispagestyle{empty}
\pagestyle{empty}
\begin{abstract}
	This paper presents a parallel Monte Carlo simulation based performance quantification method for nonlinear model predictive control (NMPC) in closed-loop. The method provides distributions for the controller performance in stochastic systems enabling performance quantification. We perform high-performance Monte Carlo simulations in C enabled by a new thread-safe NMPC implementation in combination with an existing high-performance Monte Carlo simulation toolbox in C. We express the NMPC regulator as an optimal control problem (OCP), which we solve with the new thread-safe sequential quadratic programming software NLPSQP. Our results show almost linear scale-up for the NMPC closed-loop on a 32 core CPU. In particular, we get approximately 27 times speed-up on 32 cores. We demonstrate the performance quantification method on a simple continuous stirred tank reactor (CSTR), where we perform 30,000 closed-loop simulations with both an NMPC and a reference proportional-integral (PI) controller. Performance quantification of the stochastic closed-loop system shows that the NMPC outperforms the PI controller in both mean and variance.

\end{abstract}

\section{Introduction}

In closed-loop systems, there exist many unknown or uncertain quantities, such as parameters, measurement noise, and process noise, even when simple linear controllers like proportional–integral–derivative (PID) controllers are applied. As such, achieving useful closed-loop performance quantification can be difficult. Previous work has focused on development of a high-performance Monte Carlo simulation toolbox for parallel computing on shared memory architectures. The Monte Carlo simulation toolbox has previously enabled tuning of PID controllers in closed-loop systems \cite{wahlgreen:2021a}, tuning of a model predictive controller (MPC) through controller matching \cite{wahlgreen:2023a}, and been applied for PID closed-loop insulin dosing in a virtual clinical trial with $1,000,000$ participants for people with type 1 diabetes \cite{reenberg:2022a}. Similar results have not yet been obtained for more advanced controllers such as nonlinear MPC (NMPC).

Monte Carlo approaches have previously been applied in relation to NMPC. Sequential Monte Carlo (SMC) has been applied as a method to find global optimizers in NMPC \cite{kantas:2009a}. Additionally, an SMC filter has been applied as an alternative to Kalman filtering and moving horizon estimation (MHE) for state estimation \cite{botchu:2007a}. However, application of Monte Carlo simulation to quantify the closed-loop performance of NMPC is novel, likely due to the difficulty of running sufficiently many closed-loop simulations with NMPC. We propose to apply the existing Monte Carlo simulation toolbox as a performance quantification technique for NMPC closed-loop systems. Computational feasibility is achieved by full utilization of multi-core CPUs \cite{ross:2008a}. To this end, the approach requires a thread-safe NMPC implementation.

In this paper, we apply parallel Monte Carlo simulation as a method for performance quantification of NMPC in closed-loop. The method provides performance distributions of the stochastic closed-loop and enables performance quantification. We achieve parallel scaling by implementation of a new thread-safe NMPC featuring a continuous-discrete extended Kalman filter (CD-EKF) for state estimation and a regulator expressed as an optimal control problem (OCP). We solve the OCP with a new thread-safe sequential quadratic programming (SQP) software, NLPSQP (nonlinear-programming-sequential-quadratic-programming), required to achieve parallel scaling. Due to thread-safety of the NMPC, we achieve almost linear parallel scaling and approximately 27 times speed-up on 32 cores. The efficient parallel framework enables computationally feasible Monte Carlo simulations of closed-loop systems with NMPC, which enables novel performance quantification of NMPC. We consider a well-known continuous stirred tank reactor (CSTR) example as a case study \cite{wahlgreen:2020a,jorgensen:2020a}. We demonstrate the performance quantification method by comparing NMPC performance to a reference proportional-integral (PI) controller. The resulting performance distributions show that the NMPC outperforms the PI controller in both mean and variance.

The remaining parts of the paper are organized as follows. Section \ref{sec:CDS} introduces the continuous-discrete system. Section \ref{sec:NMPC} introduces our NMPC formulation including the estimator and regulator. Section \ref{sec:NLPSQP} presents the SQP software NLPSQP. Section \ref{sec:PI} presents the PI controller. Section \ref{sec:model} introduces our case study model. Section \ref{sec:results} presents our results. Section \ref{sec:Conclusion} presents our conclusions.

\section{Continuous-discrete System}
\label{sec:CDS}

In our closed-loop simulations, we consider stochastic continuous-discrete systems in the form \cite{wahlgreen:2021a},
\begin{subequations} \label{eq:CDS}
	\begin{alignat}{3}
		\begin{split}
			d x(t) &= f(t,x(t),u(t),d(t),p) dt 
			\\ & \qquad + \sigma(t,x(t),u(t),d(t),p) d\omega(t), \label{eq:CDS_states}
		\end{split} \\
		y(t_i) &= g(t_i,x(t_i),p) + v(t_i,p), \\
		z(t) &= h(t,x(t),p),
	\end{alignat}	
\end{subequations}
where $x(t)$ are states, $u(t)$ are inputs, $d(t)$ are disturbances, $p$ are parameters, $y(t_i)$ are measurements at discrete time, $z(t)$ are outputs, $\omega(t)$ is a standard Wiener process, and $v(t_i,p)$ is normally distributed measurement noise at discrete time, i.e.,
\begin{subequations}
	\begin{align}
		d\omega(t) &\sim N_{iid}(0,I dt), \\
		v(t_i, p) &\sim N_{iid}(0,R(t_i,p)),
	\end{align}
\end{subequations}
where $R$ is the measurement covariance. Measurements, $y(t_i)$, are assumed available with sampling time, $T_s$. We apply models in the stochastic continuous-discrete form, (\ref{eq:CDS}), for both simulation of the system and in the NMPC. 


\section{Nonlinear model predictive controller}
\label{sec:NMPC}

We design an NMPC scheme to regulate continuous-discrete systems in the form (\ref{eq:CDS}). Our NMPC includes a CD-EKF for state estimation \cite{wahlgreen:2021a, simon:2016a} and a regulator expressed as an OCP.

\subsection{State estimator}
The CD-EKF receives a measurement, $y_i$, at time $t_i$. It computes the state-covariance one-step prediction, $\hat x_{i|i-1}$ and $P_{i|i-i}$, from the previous state-covariance estimate, $\hat x_{i-1|i-1}$ and $P_{i-i|i-1}$, and applies the measurement and the one-step prediction to compute the new filtered state-covariance estimate, $\hat x_{i|i}$ and $P_{i|i}$. 

\subsubsection{Prediction}
The state and covariance one-step prediction is,  
\begin{align}
	\hat x_{i|i-1} &= \hat x_{i-1}(t_i), & P_{i|i-1} &= P_{i-1}(t_i),
\end{align}
obtained as the solution to, 
\begin{subequations}
	\label{eq:predictionEquations}
	\begin{alignat}{3}
		\frac{d}{dt} \hat x_{i-1}(t) &= f(t,\hat x_{i-1}(t),u_{i-1},d_{i-1}, p), \\
		\begin{split}
			\frac{d}{dt} P_{i-1}(t) &= A_{i-1}(t) P_{i-1}(t) + P_{i-1}(t) A_{i-1}(t)^{\top}
			\\ & \quad + \sigma_{i-1}(t) \sigma_{i-1}(t)^{\top},
		\end{split}
	\end{alignat}
\end{subequations}
for $t_{i-1} \leq t \leq t_i$, where
\begin{subequations}
	\begin{alignat}{3}
		A_{i-1}(t) &= \frac{\partial }{\partial x} f(t,\hat x_{i-1}(t),u_{i-1},d_{i-1},p), \\
		\sigma_{i-1}(t) &= \sigma(t,\hat x_{i-1}(t),u_{i-1},d_{i-1},p).
	\end{alignat}
\end{subequations}
The initial condition of (\ref{eq:predictionEquations}) is the previous filtered state-covariance pair, 
\begin{subequations}
	\begin{align}
		\hat x_{i-1}(t_{i-1}) &= \hat x_{i-1|i-1}, & P_{i-1}(t_{i-1}) &= P_{i-1|i-1}.
	\end{align}
\end{subequations}

\subsubsection{Filtering}
Given the measurement, $y_i$, and the state-covariance one-step prediction, $\hat x_{i|i-1}$ and $P_{i|i-1}$, the CD-EKF computes the filtered state estimate, $\hat x_{i|i}$, as
\begin{subequations}
	\begin{align}
		&\hat y_{i|i-1} = g(\hat x_{i|i-1},p), &  &C_i = \frac{\partial}{\partial x}g(\hat x_{i|i-1},p), \\
		&e_i = y_i - \hat y_{i|i-1}, &  &R_{e,i} = R_i + C_i P_{i|i-1}C_i^{\top}, \\
		&\hat x_{i|i} = \hat x_{i|i-1}+K_ie_i, &  &K_i = P_{i|i-1}C_i^{\top}R_{e,i}^{-1},
	\end{align}
\end{subequations}
where $R_i = R(t_i,p)$ is the measurement covariance. The filtered covariance estimate, $P_{i|i}$, is
\begin{alignat}{3}
	\begin{split}
		P_{i|i} &= P_{i|i-1} - K_{i} R_{e,i} K_{i}^{\top} \\
		&= (I - K_iC_i) P_{i|i-1} (I - K_i C_i)^{\top} + K_i R_i K_i^{\top}, \label{eq:Josph}
	\end{split}
\end{alignat}
where (\ref{eq:Josph}) is the Joseph stabilizing form \cite{schneider:2013a}.


\subsection{Regulator}
We express the NMPC regulator in terms of an OCP, which the NMPC solves at time $t_i$ once the filtered state estimate, $\hat x_{i|i}$, is provided by the estimator. The solution to the OCP is the input and state trajectories in the finite horizon. However, the regulator only implements the first input, $u_i$, and resolves the OCP once the next state estimate is available from the estimator. Let $T$ be the prediction and control horizon, which is split into $N$ control intervals of size $T_{s}$. As such, $T = N T_s$. We assume zero-order hold parameterization of inputs, $u$, and disturbances, $d$, in each control interval,
\begin{subequations}
	\begin{align}
		u(t) &= u_{i+k}, && t_{i+k} \leq t < t_{i+k+1}, \\
		d(t) &= d_{i+k}, && t_{i+k} \leq t < t_{i+k+1},
	\end{align}
\end{subequations}
where $t_{i+k} = t_i + kT_s$. Let $\mathcal{N} = \{0,1,..., N-1\}$, then the regulator OCP is, 
\begin{subequations}
	\label{eq:OPC_general}
	\begin{alignat}{3}
		& \min_{x,u} \quad && \varphi_i, \\
		& \hspace{0.2cm} \text{s.t.} && x(t_i) = \hat x_{i|i}, \\
		& && \dot x(t) = f(t,x,u,d,p), \hspace{0.4cm}&& t_i \leq t \leq t_{i}+T, \\
		& && u(t) = u_{i+k}, \quad k \in \mathcal{N}, &&  t_{i+k} \leq t \leq t_{i+k+1},  \\
		& && d(t) = d_{i+k}, \quad k \in \mathcal{N}, &&  t_{i+k} \leq t \leq t_{i+k+1},  \\
		& && u_{\min} \leq u_{i+k} \leq u_{\max}, && k \in \mathcal{N}, 
	\end{alignat}
\end{subequations}
where $f(t,x,u,d,p) = f(t,x(t),u(t),d(t),p)$ and $\varphi_i = \varphi_i( x(t), u(t) )$.
We apply a direct multiple-shooting discretization to solve the OCP, (\ref{eq:OPC_general}), which yields a nonlinear programming (NLP) in the form,
\begin{subequations}
	\label{eq:OCP_NLP}
	\begin{align}
		\min_{ \xi } \quad & \varphi, \\
		\text{s.t.} \quad & x_{k+1} - F(t_{k}, x_{k}, u_{k}, d_{k}, p) = 0, \label{eq:OCP_NLP_eqCons} \\
		& u_{\min} \leq u_{k} \leq u_{\max}, \label{eq:OCP_NLP_bound}
	\end{align}
\end{subequations}
where $k \in \mathcal{N}$ is relative in time to $t_i$, $x_0 = \hat x_{0|0}$ is a parameter, $F(\cdot)$ is a numerical state integration scheme, and the decision variables are,
\begin{align}
	\xi_i = \begin{bmatrix}
		u_{0} &
		x_{1} &
		\cdots 	&
		u_{N-1} &
		x_{N}
	\end{bmatrix}^{\top}.
\end{align}
Note that $x_0$ is not required as a decision variable, but can be included without loss of generality. We assume that the NLP objective, $\varphi = \varphi(\xi)$, is partially separable locally in time with respect to the decision variables. Additionally, we denote the number of equality constraints, $m_e$, the number of lower bounds, $m_l$, and the number of upper bounds, $m_u$.


\section{Sequential Quadratic Programming}
\label{sec:NLPSQP}

We solve the NLP, (\ref{eq:OCP_NLP}), with our SQP software, NLPSQP. The NLPSQP implementation is dedicated to solve multiple similar NLPs in parallel applications. The iterative SQP algorithm performs three steps in each iteration, 1) Obtain a search direction by solution of a Quadratic Programming (QP) subproblem, 2) Obtain a step-size by a backtracking line-search algorithm, and 3) Lagrangian Hessian approximation with a block Broyden–Fletcher–Goldfarb–Shanno (BFGS) update. 

In the following, we let $g(\xi)$ denote the vectorized constraint evaluation of (\ref{eq:OCP_NLP_eqCons}) together with $\lambda$, $\pi_l$, and $\pi_u$ denoting Lagrange multipliers for equality constraint, lower input bound constraints, and upper input bound constraints respectively. Additionally, we let $f(\xi) = \varphi(\xi)$ for simplicity and apply $[l]$ as superscript to denote the $l$'th iteration of the algorithm.

\subsection{Quadratic Programming subproblem}
In iteration $l$, the QP-subproblem solved in NLPSQP is
\begin{subequations}
	\label{eq:QP-subproblem}
	\begin{align}
		\min_{  \Delta \xi  } \quad &\bar l_0(\Delta u_0) + \sum_{k=1}^{N-1} \bar l_k(\Delta x_k, \Delta u_k) + \bar l_N(\Delta x_N), \\
		\text{s.t.} \quad &\Delta x_{k+1} = A_k^{\top} \Delta x_k + B_k^{\top} \Delta u_k + b_k,  \\
		& u_{\min} - u_k \leq \Delta u_k \leq u_{\max} - u_k, 
	\end{align}
	\label{eq:OCP}
\end{subequations}
where $k \in \mathcal{N}$, $\Delta x_0 = 0$ is a parameter, and
\begin{subequations}
	\label{eq:nlpsqp_qp}
	\begin{align}
		\bar l_0(\Delta u_0) ={} \frac{1}{2}&\Delta u_0^{\top} R_0 \Delta u_0 + r_0^{\top} \Delta u_0 + \rho_0, \\
		\begin{split}
			\bar l_k(\Delta x_k, \Delta u_k) ={} \frac{1}{2} &\begin{bmatrix}
				\Delta x_k \\
				\Delta u_k
			\end{bmatrix}^{\top} \begin{bmatrix}
				Q_k & M_k \\
				M_k^{\top} & R_k	
			\end{bmatrix} \begin{bmatrix}
				\Delta x_k \\
				\Delta u_k
			\end{bmatrix}  \\
			+ &\begin{bmatrix}
				q_k \\
				r_k
			\end{bmatrix}^{\top} \begin{bmatrix}
				\Delta x_k \\
				\Delta u_k
			\end{bmatrix} + \rho_k, 
		\end{split}\\
		\bar l_N(\Delta x_N) ={} \frac{1}{2} &\Delta x_N^{\top} P_N \Delta x_N + p_N^{\top}\Delta x_N + \rho_N.
	\end{align}
\end{subequations}
Due to partial separability of the Lagrangian function, 
\begin{align}
	\begin{split}
		&\mathcal{L}( \xi, \lambda, \pi_l, \pi_u ) = \mathcal{L}_0(u_0, \lambda, \pi_l, \pi_u) \\
		&\qquad+ \sum_{k=1}^{N-1} \mathcal{L}_k (x_k, u_k, \lambda, \pi_l, \pi_u) + \mathcal{L}_N(x_N, \lambda),
	\end{split}
\end{align}
the Lagrangian Hessian is block diagonal with blocks, $W_k$, defined as,
\begin{align}
	W_0 &= R_0, &
	W_k &=  \begin{bmatrix}
		Q_k 		& M_k \\
		M_k^{\top}	& R_k
	\end{bmatrix}, &
	W_N &= P_N,
\end{align}
for $k = 1,...,N-1$. The matrices, $Q_k$, $R_k$, $M_k$, and $P_N$, are second order derivatives of the Lagrangian function. However, NLPSQP applies a BFGS type approximation for the blocks, $W_k$ \cite{bock:1984a}. The remaining matrices and vectors in the QP-subproblem, (\ref{eq:QP-subproblem}), are given as,
\begin{subequations}
	\begin{align}
		r_k &= \nabla_{u_k} \mathcal{L}_k,	&& k = 0,..., N-1, 	\\
		q_k &= \nabla_{x_k} \mathcal{L}_k, 	&& k = 1,..., N-1,	\\
		p_N &= \nabla_{x_N} \mathcal{L}_N, 						\\
		A_k &= \nabla_{x_k} F_k,			&& k = 1,..., N-1,	\\
		B_k &= \nabla_{u_k} F_k,			&& k = 0,..., N-1,	\\
		b_k &= F_k - x_{k+1},				&& k = 0,..., N-1,
	\end{align}
\end{subequations}
where $F_k = F(t_k, x_k, u_k, d_k, p)$. The solution to the QP-subproblem, (\ref{eq:nlpsqp_qp}), is the data $(\Delta \xi, \mu, \nu_l, \nu_u)^{[l]}$, where 
\begin{subequations}
	\begin{align}
		\mu^{[l]} 	&= \lambda^{[l]} + \Delta \lambda^{[l]}, 	\\
		\nu_l^{[l]} &= \pi_l^{[l]} + \Delta \pi_l^{[l]}, 		\\
		\nu_u^{[l]} &= \pi_u^{[l]} + \Delta \pi_u^{[l]}.
	\end{align}
\end{subequations}
Notice that the search direction, $(\Delta \xi, \Delta \lambda, \Delta \pi_l, \Delta \pi_u)^{[l]}$, ensures satisfaction of the linear bound constraints, (\ref{eq:OCP_NLP_bound}), if the initial guess is feasible. NLPSQP solves the structured QP-subproblem, (\ref{eq:QP-subproblem}), with a Riccati recursion based primal-dual interior point algorithm \cite{frison:2013a,jorgensen:2004a,wahlgreen:2022a}. 



\subsection{Line-search}
Given the search direction, $(\Delta \xi, \Delta \lambda, \Delta \pi_l, \Delta \pi_u)^{[l]}$, NLPSQP performs the step,
\begin{align}
	\begin{split}
		(\xi, \lambda, \pi_l, \pi_u)^{[l+1]} &= (\xi, \lambda, \pi_l, \pi_u)^{[l]} \\
		&\quad + \alpha (\Delta \xi, \Delta \lambda, \Delta \pi_l, \Delta \pi_u)^{[l]},
	\end{split}
\end{align}
where $\alpha$ is a step-size. NLPSQP applies a backtracking line-search algorithm to select a step-size, $\alpha$, ensuring sufficient decrease in Powell's $l_1$-merit function \cite{jorgensen:2004a,powell:1978a},
\begin{align}
	P(\xi, \sigma) = f(\xi) + \sigma^{\top} |g(\xi)|,
\end{align}
where
\begin{align}
	\sigma_i = \max\left( |\mu_i|, \frac{1}{2}(\sigma_i + |\mu_i|) \right), \quad i = 1,...,m,
\end{align}
with $\sigma_i = |\mu_i|$ in the first iteration. We define
\begin{align} 
	T(\alpha) = P(\xi^{[l+1]}, \sigma) = P(\xi^{[l]} + \alpha \Delta \xi^{[l]}, \sigma),
\end{align}
and let sufficient decrease be defined from the Armijo condition,
\begin{align} \label{eq:Armijo}
	T(\alpha) \leq T(0) + c_1 \alpha \mathrm{D}_{\Delta \xi}T(0),
\end{align}
where
\begin{align}
	T(\alpha) &= f(\xi^{[l]} + \alpha \Delta \xi^{[l]}) + \sigma^{\top}|g(\xi^{[l]} + \alpha\Delta \xi^{[l]})|,  \\
	T(0) &= f(\xi^{[l]}) + \sigma^{\top}|g(\xi^{[l]})|, \\
	\mathrm{D}_{\Delta \xi}T(0) &= \nabla f(\xi^{[l]})^{\top}\Delta \xi^{[l]} - \sigma^{\top} |g(\xi^{[l]})|.
\end{align}
The backtracking line-search algorithm is, 
\begin{enumerate}
	\item Set $\alpha = 1$
	\item\label{enum:2} Evaluate (\ref{eq:Armijo}). If satisfied, \textbf{break} with $\alpha$ as output
	\item Compute $\alpha = \beta \alpha$
	\item Go to \ref{enum:2})
\end{enumerate}
where $0 < \beta < 1$. We use $c_1 = 10^{-4}$ and $\beta = 0.5$ (similarly to IPOPT \cite{waechter:2006a}).

\subsection{Block BFGS update}
NLPSQP estimates the block matrices, $W_k$, with a block damped BFGS update \cite{bock:1984a}. Define
\begin{subequations} \label{eq:sy}
	\begin{align}
		s &= \xi^{[l+1]} - \xi^{[l]}, \\
		y &= \nabla_{\xi} \mathcal{L}^{[l+1]} - \nabla_{\xi} \mathcal{L}^{[l]},
	\end{align}
\end{subequations}
where $\nabla_{\xi} \mathcal{L}^{[l]} = \nabla_{\xi} \mathcal{L}(\xi^{[l]}, \lambda^{[l+1]}, \pi_{l}^{[l+1]}, \pi_u^{[l+1]})$, $\nabla_{\xi} \mathcal{L}^{[l+1]} = \nabla_{\xi} \mathcal{L}(\xi^{[l+1]}, \lambda^{[l+1]}, \pi_{l}^{[l+1]}, \pi_u^{[l+1]})$, and let $s_k$ and $y_k$ be the elements of $s$ and $y$ corresponding to $W_k$, respectively. Let
\begin{align}
	r_k = \theta_k y_k + (1-\theta_k)W_k s_k,
\end{align}
where
\begin{align}
	\theta_k = \begin{cases}
		1 & s_k^{\top} y_k \geq 0.2s_k^{\top} W_k s_k \\
		\frac{0.8 s_k^{\top} W_k s_k}{s_k^{\top}W_k s_k - s_k^{\top} y_k} & \text{else}
	\end{cases}
\end{align}
Then the block BFGS update is given by
\begin{align}
	W_{k+1} = \begin{cases}
		W_k - \frac{(W_k s_k)(W_k s_k)^{\top}}{s_k^{\top} (W_k s_k)} + \frac{r_k r_k^{\top}}{s_k^{\top} r_k} & \kappa > \epsilon_m \\
		W_k & \text{else}
	\end{cases}
\end{align}
where $\epsilon_m$ is the machine precision, $\kappa = \min(\kappa_1, \kappa_2)$ with $\kappa_1 = s_k^{\top}W_ks_k$ and $\kappa_2 = s_k^{\top} r_k$. The update safeguard is required as some blocks might converge faster than others resulting in zero-division. NLPSQP initializes the Hessian update as identity, $W^{[0]} = I$. Numerical rounding errors might cause indefinite BFGS block updates. In this case, NLPSQP applies the simple strategy to reset the entire Hessian to identity. 


\subsection{Convergence}
NLPSQP converges when the KKT conditions are satisfied to the user-specified tolerance $\epsilon$. In practice, we apply a scaled convergence condition, 
\begin{align} \label{eq:convergence}
	|| \nabla \mathcal{L}^{[l]}/s_d ||_{\infty} &\leq \epsilon, &
	||g^{[l]}||_{\infty} &\leq \epsilon,
\end{align}
where $\nabla \mathcal{L}^{[l]} = \nabla \mathcal{L}(\xi, \lambda, \pi_l, \pi_u)^{[l]}$, $g^{[l]} = g(\xi)^{[l]}$, and
\begin{align}
	s_d = \max\left( s_{\max}, \frac{|| \lambda ||_1 + ||\pi_l||_1 + ||\pi_u||_1}{m_e + m_l + m_u} \right)/s_{\max},
\end{align}
with $s_{\max} = 100$ (similar to IPOPT \cite{waechter:2006a}).

\subsection{Implementation}
NLPSQP is implemented thread-safe in C for parallel applications, specifically intended for closed-loop Monte Carlo simulation. NLPSQP is BLAS dependent. In particular, we apply OpenBLAS \cite{xianyi:2012a,wang:2013a}. To ensure thread-safety of OpenBLAS, we compile a single threaded version by setting \texttt{USE\_THREAD=0} and \texttt{USE\_LOCKING=1}. However, we have not been able to achieve parallel scaling for the functions \texttt{dpotrf}, \texttt{dpotrs}, and \texttt{dgemm}. Therefore, we have implemented our own versions of these functions only intended for small matrices. In future work, we will consider other thread-safe BLAS libraries.


\section{Proportional–integral controller}
\label{sec:PI}

We consider a PI controller with input bounds, $u_{\min}$ and $u_{\max}$, and anti-windup mechanism to ensure proper integrator behavior when the PI response is saturated,
\begin{subequations} \label{eq:PI}
	\begin{align}
		e_k &= \bar y_k - y_k, \\
		P_k &= k_P e_k, \\
		I_k &= \hat I_{k-1} + T_s k_I e_k, \\
		\hat u_k &= \bar u + P_k + I_k, \\
		u_k &= \max( u_{\text{min}}, \min( u_{\max}, \hat u_k ) ), \label{eq:PI_clip} \\
		I_{aw,k} &= T_s k_{aw}( \hat u_k - u_k ), \label{eq:PI_aw} \\
		\hat I_{k} &= I_k + I_{aw, k} \label{eq:PI_Ihat}.
	\end{align}
\end{subequations}
We apply the PI controller (\ref{eq:PI}) as a reference for the NMPC.

\section{Model}
\label{sec:model}

As a simulation case study, we consider an exothermic chemical reaction conducted in an adiabatic CSTR \cite{wahlgreen:2020a,jorgensen:2020a}. The example is simple yet effective due to the non-trivial dynamics causing a branch of unstable steady-states in the operating window \cite{wahlgreen:2020a}. In addition, the model is well-approximated by a one-state model, well-suited for NMPC. The stochastic model for the constant volume CSTR is compactly written as the SDE \cite{wahlgreen:2023a},
\begin{align} \label{eq:CSTR}
	dn(t) = \left( C_{\mathrm{in}}F - c F + RV \right) dt + F \bar \sigma d\omega(t),
\end{align}
where
\begin{align}
	&c = \frac{n}{V}, && R = S^{\top}r(c),
\end{align}
and
\begin{align}
	&r(c) = k(c_T) c_A c_B, && k(c_T) = k_0 \exp\left( -\frac{E_a}{R}\frac{1}{c_T} \right),
\end{align}
with $E_a/R$ denoting the activation energy. In the three-state model, the stoichiometric matrix and inlet stream concentration matrix is,
\begin{align}
	&C_{\mathrm{in}} = \begin{bmatrix} 
		c_{A,\mathrm{in}} \\
		c_{B,\mathrm{in}} \\
		c_{T,\mathrm{in}} 
	\end{bmatrix}, && S = \begin{bmatrix}
		-1.0 & -2.0 & \beta
	\end{bmatrix},
\end{align} 
and in the one-state model they are,
\begin{align}
	C_{in} = \begin{bmatrix}
		c_{T,\mathrm{in}}
	\end{bmatrix}, \quad S = \begin{bmatrix}
		\beta
	\end{bmatrix}.
\end{align}
The stochastic diffusion term of (\ref{eq:CSTR}) models inlet concentration variations and we apply
\begin{align}
	\bar \sigma &= \begin{bmatrix}
		\sigma_A \\
					& \sigma_B \\
					&			& \sigma_T
	\end{bmatrix}, & \bar \sigma = \begin{bmatrix}
		\sigma_T
	\end{bmatrix},
\end{align}
in the three-state and one-state models, respectively. We refer to \cite{wahlgreen:2020a} for more details and the parameters of the model. The output of the model is the temperature $c_T$, i.e., $z(t) = c_T(t)$. Additionally, we assume the temperature to be measured at discrete times, i.e., $y(t_i) = c_T(t_i)$.

\section{Results}
\label{sec:results}

This section presents our simulation results. The simulations are conducted on a dual-socket Intel(R) Xeon(R) Gold 6226R CPU @ 2.90GHz system. See TABLE \ref{tab:cpu_details} for CPU details. 
\begin{table}[tb]
	\centering
	\caption{CPU information.}
	\label{tab:cpu_details}
	\begin{tabular}{l|l}
		Architecture: & x86\_64  \\
		CPU op-mode(s): & 32-bit, 64-bit \\
		CPU(s): & 32 \\
		Thread(s) per core: & 1 \\
		Core(s) per socket: & 16 \\
		Socket(s): & 2 \\
		NUMA node(s): & 2 \\
		Model name: & Intel(R) Xeon(R) Gold 6226R CPU @ 2.90GHz \\
		CPU MHz: & 2900.000 \\
		L1d cache: & 32 kB \\
		L1i cache: & 32 kB \\
		L2 cache: & 1024 kB \\
		L3 cache: & 22528 kB \\
		RAM: & 384 GB
	\end{tabular}
\end{table}

\subsection{Closed-loop simulation}

\begin{figure}
	\centering
	\includegraphics[width=0.48\textwidth]{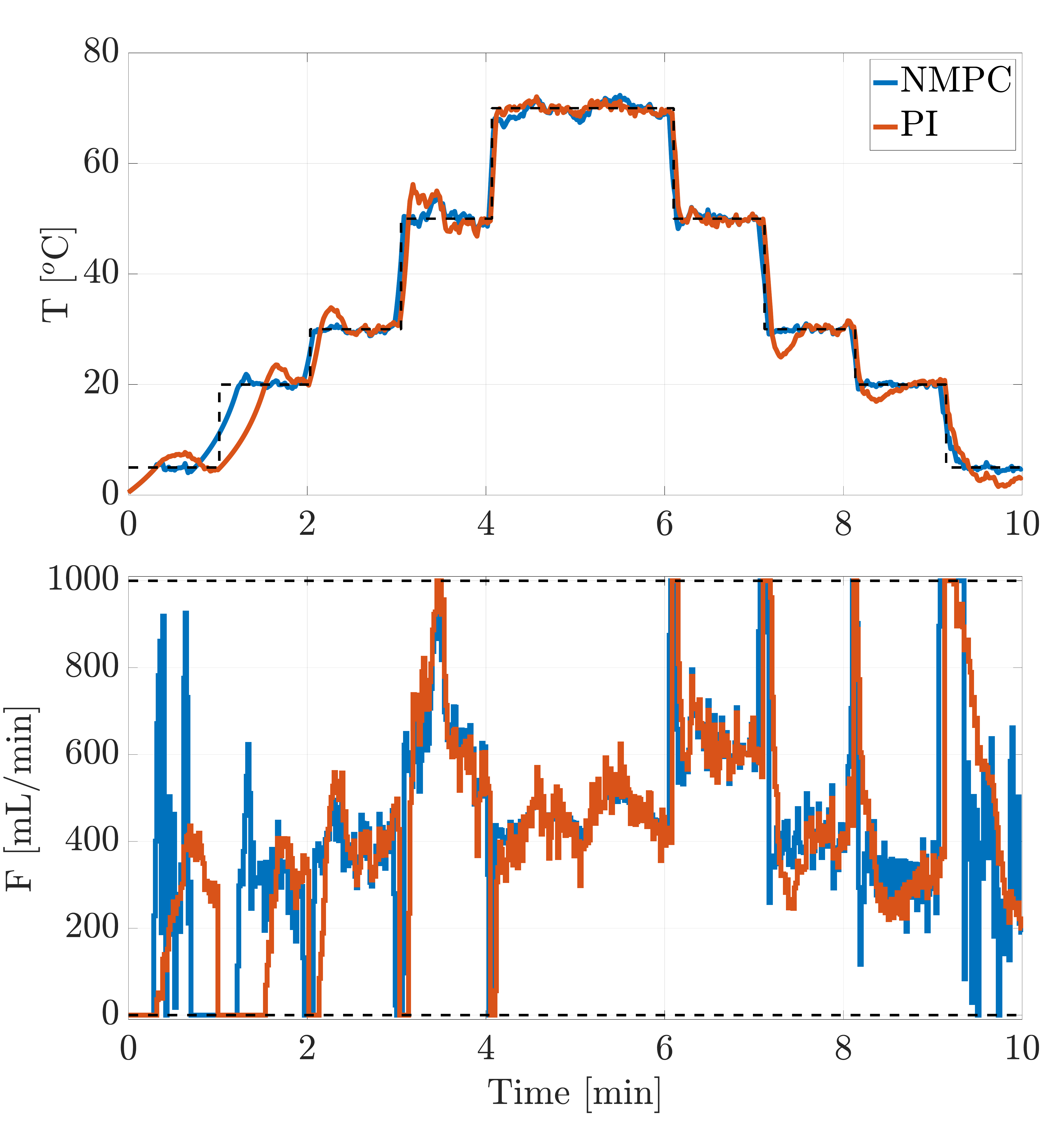}
	\caption{ Stochastic closed-loop simulation with PI controller and NMPC. Both controllers are able to track the set-points at both stable and unstable steady-states. The NMPC has better tracking performance at set-point changes due to its anticipatory action. }
	\label{fig:closedloopsimulationf1}
\end{figure}

We simulate the CSTR in closed-loop with the NMPC. We apply the three-state stochastic model for simulation of the system and apply the one-state model in the NMPC. We select a variable set-point, $\bar z$, together with, $t_0 = 0.0$ s, $t_f = 600.0$ s, $T_s = 1.0$ s, and $N_s = 20$, where $N_s$ is the number of Euler-Maruyama steps to integrate the state equations from $t_i$ to $t_{i+1}$. We initialize the system at 
\begin{align}
	x_0 = n_0 = \begin{bmatrix} 
		c_{A,\mathrm{in}} \\
		c_{B,\mathrm{in}} \\
		c_{T,\mathrm{in}}
	\end{bmatrix}V.
\end{align}
The NMPC has the discrete prediction and control horizon, $N = 60$, applies $N_c = 5$ classical Runge-Kutta steps to integrate the state dynamics in each control interval in the OCP, and applies a point-wise weighted least-squares output objective in the OCP,
\begin{align}
	\varphi_{i} = \sum_{k=1}^{N} ||z( t_{i+k} ) - \bar z( t_{i+k} ) ||_{Q_z}^2 T_s,
\end{align}
where $Q_z = 1.0$. We initialize the CD-EKF states as,
\begin{align}
	&\hat x_{-1|-1} = c_{T, \mathrm{in}}V, && P_{-1|-1} = 10^{-6},
\end{align}
and apply the input bounds $u_{\min} = 0.0$ mL/min and $u_{\max} = 1000.0$ mL/min. The PI controller with anti-windup mechanism, (\ref{eq:PI}), has the hand-tuned gains,
\begin{subequations}
	\begin{align}
		k_P &= -10^{-3}, &k_I &= -10^{-4}, &k_{aw} &= -10^{-1}.
	\end{align}
\end{subequations}
We perform a single closed-loop simulation with the PI controller and the NPMC. Fig. \ref{fig:closedloopsimulationf1} presents the result. We observe that both the PI controller and the NMPC are able to track set-points at both stable and unstable steady-states. However, the NMPC has better tracking performance at set-point changes due to its anticipatory action.

\begin{table}
	\centering
	\caption{Statistics for Monte Carlo simulation of NMPC closed-loop. }
	\label{tab:MC_stat}
	\begin{tabular}{l|r l}			
		MC simulation time 				& $\approx$ 55 	& [min]	\\
		Number of MC simulations 		& 30,000		& [-]	\\
		Total number of OCPs			& 18,000,000	& [-]	\\
		Successful OCPs					& 17,999,576	& [-]	\\
		Failed OCPs						& 424 			& [-]	\\
		Percentage success				& 99.9976		& [\%]	\\
		Percentage fails				& 0.0024		& [\%]
	\end{tabular}
\end{table}

\subsection{Parallel scalability}
\begin{figure}[tb]
	\centering
	\includegraphics[width=0.48\textwidth]{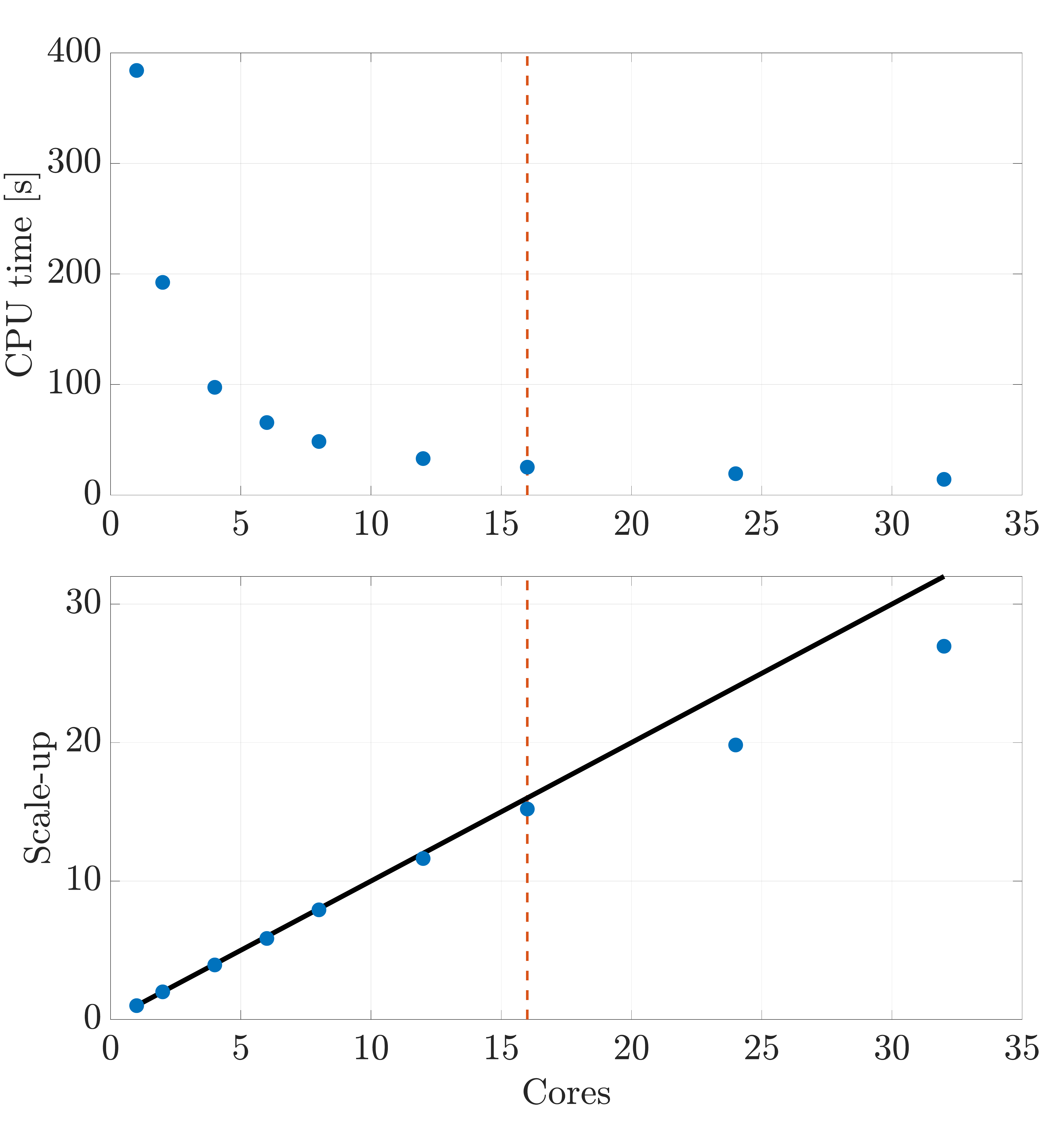}
	\caption{Scaling plot for Monte Carlo simulations of NMPC closed loop in parallel. Scaling on 32 cores is approximately 27 times.}
	\label{fig:scaleupplotf1}
\end{figure}

We apply the Monte Carlo simulation toolbox to perform 100 simulations with different process noise in the system. We compute the 100 simulations with the NMPC on different numbers of cores to get scale-up data. Fig. \ref{fig:scaleupplotf1} presents a scale-up plot for the simulations. We observe almost linear scale-up and approximately $27$ times speed-up on 32 cores. In previous work, we showed similar scale-up results for a PID controller \cite{wahlgreen:2021a}.

\subsection{Monte Carlo Simulations}

\begin{figure}[tb]
	\centering
	\includegraphics[width=0.48\textwidth]{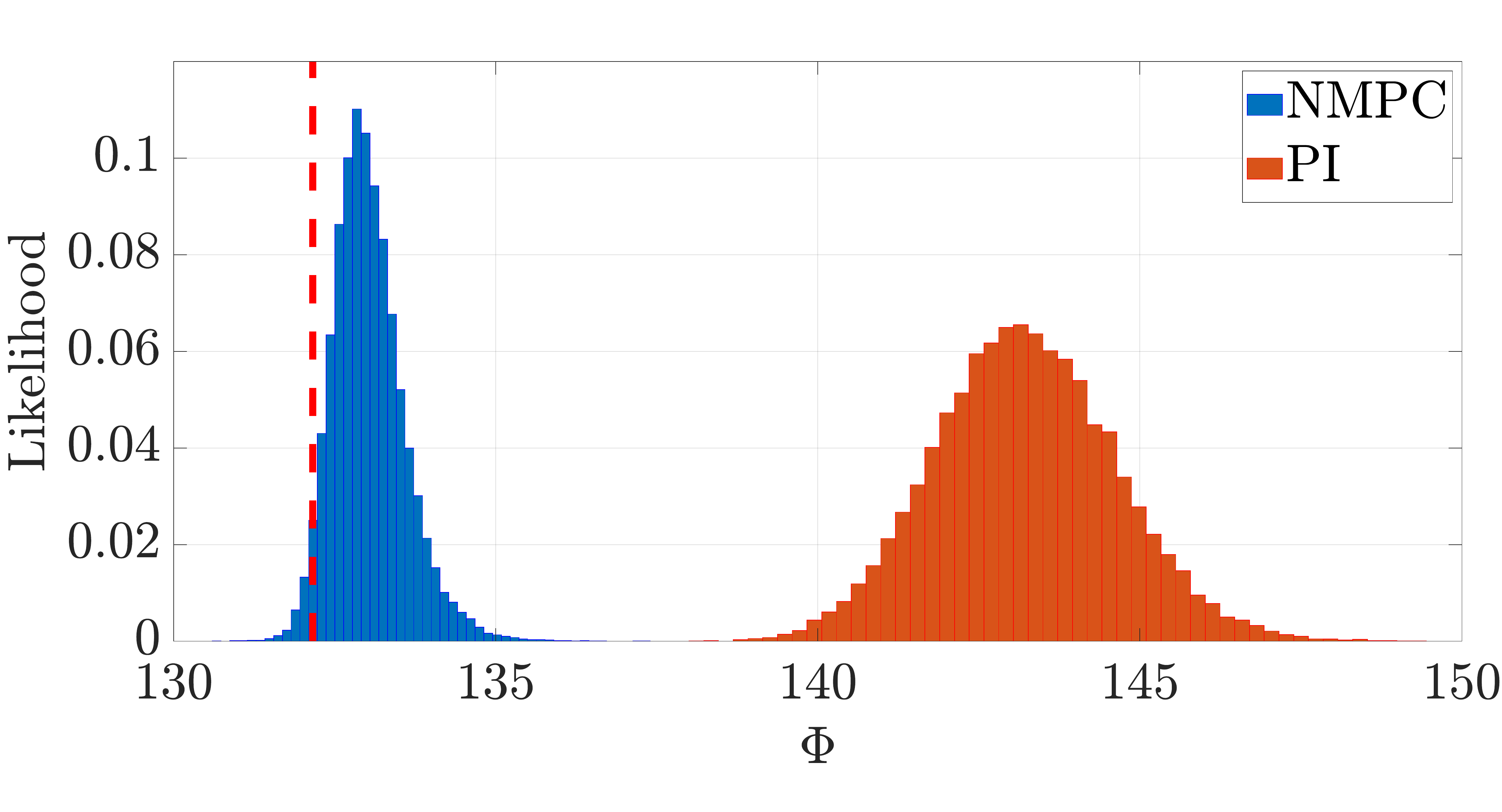}
	\caption{Histograms of the distribution of the $\Phi$-metric, (\ref{eq:Phi}), for the PI controller and the NMPC in closed-loop. The red dashed line indicates the performance of the NMPC in the deterministic case. Computation time for 30,000 closed-loop simulations with NMPC is approximately $55$ min.}
	\label{fig:pdff1}
\end{figure}

We perform Monte Carlo simulations to quantify the closed-loop performance of the NMPC in presences of process noise. In particular, we perform $30,000$ simulations with varying process noise for both the PI controller and the NMPC. We apply a scaled point-wise squared-2-norm metric to evaluate the closed-loop performance,
\begin{align} \label{eq:Phi}
	\Phi = \frac{1}{\bar N+1} \sum_{i=0}^{\bar N} || z(t_i) - \bar z(t_i) ||_2^2,
\end{align}
where $\bar N$ is the number of samplings over the full simulation, i.e., $\bar N = \frac{t_f-t_0}{T_s} = 600$ in our simulations. Fig. \ref{fig:pdff1} shows historgams of the distribution of $\Phi$, (\ref{eq:Phi}), over the 30,000 Monte Carlo simulations. The results show that the NMPC outperforms the PI controller in both mean and variance with respect to the $\Phi$-metric. However, due to Fig. \ref{fig:closedloopsimulationf1}, we expect the better NMPC performance to be mainly due to set-point changes and anticipatory action from the NMPC.

TABLE \ref{tab:MC_stat} shows simulation statistics for the NMPC closed-loop Monte Carlo simulations. We observe that NLPSQP successfully solves $99.9976$\% of the $18,000,000$ OCPs in the $30,000$ closed-loop simulations. In the remaining $0.0024$\%, NLPSQP reaches the maximum number of iterations set to $100$. We suspect that NLPSQP locates an almost optimal point, but has trouble detecting it. Therefore, we implement the detected solution in the NMPC. In future work, we will further investigate the convergence detection of NLPSQP.



\section{Conclusion}\label{sec:Conclusion}

The paper presented a parallel Monte Carlo simulation based performance quantification method for NMPC in closed-loop. The method is made computationally feasible by combining a new thread-safe NMPC implementation with an existing implementation of a high-performance Monte Carlo simulation toolbox in C. The toolbox showed almost linear scale-up for the closed-loop simulations with an NMPC. We considered a simple CSTR model to demonstrate the performance quantification method. We performed $30,000$ Monte Carlo simulations of the closed-loop with NMPC in approximately $55$ min and applied the simulations to quantify the performance of the NMPC in presence of process noise. We compared the NMPC performance to a hand-tuned PI controller. Performance distributions provided by the Monte Carlo simulations showed that the NMPC outperforms the PI controller in both mean and variance. In addition, the performance quantification method is well-suited for NMPC tuning, e.g., tuning of stage cost or constraint back-off.





\bibliographystyle{IEEEtran}
\bibliography{bib/IEEEabrv,ref/References}
\end{document}